\documentstyle[preprint,aps,floats,epsfig,prd]{revtex}

\newcommand{\be}{\begin{equation}}
\newcommand{\ee}{\end{equation}}
\newcommand{\bea}{\begin{eqnarray}}
\newcommand{\ea}{\end{eqnarray}}
\newcommand{\bml}{\begin{mathletters}}
\newcommand{\eml}{\end{mathletters}}

\begin{document}

\tighten

\preprint{DCPT-03/15}
\draft




\title{de Sitter/ Anti-de Sitter global monopoles}
\renewcommand{\thefootnote}{\fnsymbol{footnote}}
\author{Bruno Bertrand \  \ and \  \ Yves Brihaye\footnote{Yves.Brihaye@umh.ac.be}}
\address{Facult\'e des Sciences, Universit\'e de Mons-Hainaut,
 B-7000 Mons, Belgium}
\author{Betti Hartmann\footnote{Betti.Hartmann@durham.ac.uk}}
\address{Department of Mathematical Sciences, University
of Durham, Durham DH1 3LE, U.K.}
\date{\today}
\setlength{\footnotesep}{0.5\footnotesep}

\maketitle
\begin{abstract}
We consider global monopoles in asymptotic de Sitter/ Anti- de Sitter
space-time. We present
the by our numerical
analysis confirmed asymptotic behaviour of the metric and Goldstone field functions.
We find that the appearance of horizons in this model depends strongly on the sign
and value of the cosmological constant as well as on the value of the gravitational coupling.
In Anti-de Sitter (AdS) space, we find that for a fixed value of the cosmological
constant, global monopoles without horizons exist only up to a critical value of the gravitational
coupling.
Moreover, we observe (in contrast to another recent study)
that the introduction
of a cosmological constant can {\bf not} render a positive
mass of the global monopole.
\end{abstract}


\renewcommand{\thefootnote}{\arabic{footnote}}

\section{Introduction}
In recent years topological defects \cite{vilenkin} in asymptotic de Sitter (dS)/
Anti- de Sitter (AdS) space-time have gained renewed interest.
This is mainly due to the proposed dS/CFT \cite{strominger}, resp.
AdS/CFT \cite{adscft} correspondences.  These correspondences 
suggest a holographic duality between gravity in a $d$-dimensional
dS, resp. AdS space and a conformal field theory (CFT) ``living''
on the boundary of the dS, resp. AdS spacetime and thus being $d-1$-dimensional. 
However, dS space-time is also interesting from a cosmological point of
view since it seems to be confirmed by observational data \cite{super}
that we live in a universe with positive cosmological constant.

Gravitating global monopoles in asymptotically flat space-time 
were first discussed in \cite{vile,harari}. These  topological defects
were found to have a negative mass and a deficit angle depending on the
vacuum expectation value (vev) of the scalar Goldstone field and the gravitational
coupling. For sufficiently high enough values of the vev the solutions
have an horizon \cite{lieb}. These solutions were named (after their string counterparts 
\cite{laguna}) ``supermassive monopoles''. 

In  \cite{li}, it was found that the introduction of a cosmological constant
can render the mass of the monopole positive. This was demonstrated by a figure showing the mass
function as function of the radial coordinate for different choices of the cosmological
constant. In our recent work, we are mainly interested in composite monopole
defects \cite{com} in dS/ AdS space-time \cite{ybe}. This is why we cross-checked the
results of \cite{li} and found discrepancies between our results and those in \cite{li}.

The paper is organised as follows: we give the model in Section II and discuss the
asymptotic behaviour, which should be compared to that in \cite{li}, in Section III.
We give our numerical results in Section IV and conclude in Section V.

\section{The Model}
We consider the following action~:
\begin{equation}
\label{action}
S =\int \left( \frac{1}{16\pi G}(R- 2 \Lambda)
- \frac{1}{2}\partial_{\mu} \xi^a \partial^{\mu} \xi^a  - 
\frac{ \lambda}{4}(\xi^a\xi^a- \eta^2)^2   \right) \sqrt{-g} d^4 x
\end{equation}
which describes a Goldstone triplet $\xi^a$, $a=1,2,3$, interacting with
gravity in an asymptotically de Sitter (dS) (for the cosmological constant
$\Lambda > 0$), resp. Anti-de Sitter (AdS) ($\Lambda < 0$) space-time.
$G$ is Newton's constant,
$\lambda$ is the self-coupling constant of the Goldstone field
 and $\eta$ the vacuum expectation value (vev) of the Goldstone field.

For the metric, the spherically symmetric Ansatz
in Schwarzschild-like coordinates reads~:
\begin{equation}
ds^{2}=g_{\mu\nu}dx^{\mu}dx^{\nu}=
-A^{2}(r)N(r)dt^2+N^{-1}(r)dr^2+r^2 (d\theta^2+\sin^2\theta
d^2\varphi)
\label{metric}
\ , \end{equation}
while for the Goldstone field, we choose the hedgehog Ansatz \cite{vile}~:
\begin{equation} 
 {\xi}^a = \eta  h(r) {e_r}^a 
\ . 
\end{equation}
We introduce the following dimensionless variable and coupling constants~:
\begin{equation}
     x = \eta r \quad , \quad 
     \alpha^2 = 4 \pi G \eta^2 \quad , \quad  
     \gamma = \frac{\Lambda}{\eta^2} \ .
\end{equation}

Varying (\ref{action}) with respect to the metric fields gives 
the Einstein equations which can be combined to give two first order
differential equations for $A$ and $\mu$:~
\begin{equation}
\label{a}
A' = \alpha^2 A x  (h')^2  
\end{equation}
\begin{equation}
\label{mu}
\mu' = \alpha^2  \left( h^2-1 + x^2\frac{\lambda}{4}(h^2-1)^2 +\frac{1}{2}
x^2N (h')^2 \right) 
\end{equation}
and $N$ and $\mu$ are related as follows:~
\begin{equation} 
\label{nmu}
N(x) = 1 - 2\alpha^2 - 2 \frac{\mu(x)}{x} 
                - \frac{\gamma}{3} x^2  \ . \
\end{equation}
Note that for $\gamma=0$, the existence of solutions without horizon
is restricted by $\alpha < \sqrt{\frac{1}{2}}$ \cite{lieb}.

Variation with respect to the matter fields yields the Euler-Lagrange
equations for the Goldstone field~:
\begin{equation}
(x^2 A N h')' = A( 2 h + \lambda x^2 h(h^2-1)) \ , \\
\label{feq}
\end{equation}
The prime denotes the derivative with respect to $x$.
Note that the equations have the same structure as  for the 
asymptotically flat space-time \cite{vile,harari}. The cosmological
constant just appears in the relation defining $\mu(x)$ and $N(x)$.

In order to solve the system of equations uniquely,
we have to introduce $4$ boundary conditions, which we choose to be~: 
\begin{equation}
     \mu(0) = 0 \quad ,  \quad h(0) = 0 \quad , \quad  A(\infty) = 1 \quad  , 
     \quad h(\infty) = 1 \quad .
\end{equation}

The dimensionless mass of the solution is determined by the asymptotic 
value $\mu(\infty)=\mu_{\infty}$  of the function $\mu(x)$ and is given
by $\mu_{\infty}/\alpha^2$.

\section{Asymptotic behaviour}
Expanding the functions
around the origin gives~:
\begin{equation}
h( x\rightarrow 0) = c_1 x + O(x^3) \ \ , \ \
\mu(x\rightarrow 0) = - \alpha^2 x + O(x^2) \ \ , \ \
A(x\rightarrow 0) = A(0)(1 + O(x^2))
\end{equation}
where $c_1$ and $A(0)$ are free parameters to be determined numerically.
The asymptotic behaviour ($x\rightarrow\infty$) is given by~:
\begin{eqnarray}
\label{asp}
 h(x >> 1) = 1
&+& \frac{3}{(\gamma-3\lambda)} \frac{1}{x^2}
- \frac{9 (4\alpha^2(\gamma - 3\lambda)+ 9 \lambda)}
{2(2 \gamma + 3\lambda)(\gamma-3\lambda)^2} \frac{1}{x^4}    \nonumber  \\
&-& \frac{36 \mu_{\infty}}
{(5\gamma+3\lambda)(\gamma-3\lambda)}
   \frac{1}{x^5}
+ O(\frac{1}{x^6}) \ ,
\end{eqnarray}
\begin{eqnarray}
A(x >>1 ) = 1
&-& \frac{9\alpha^2}{(\gamma-3\lambda)^2} \frac{1}{x^4}
+ \frac{36\alpha^2(4\alpha^2(\gamma - 3\lambda)+ 9 \lambda)}
{(2 \gamma + 3\lambda)(\gamma-3\lambda)^3} \frac{1}{x^6}    \nonumber  \\
&+& \frac{2160 \mu_{\infty} \alpha^2}
{7(5\gamma+3\lambda)(\gamma-3\lambda)^3}
   \frac{1}{x^7}
+ O(\frac{1}{x^8})
\label{ainfty}
\end{eqnarray}
and
 \begin{eqnarray}
 \label{muinfty}
\mu(x >> 1) = \mu_{\infty}
&+& \frac{9\alpha^2\lambda}{(\gamma-3\lambda)^2} \frac{1}{x}
- \frac{6\alpha^2
(\gamma^2(2\alpha^2 + 3) -12\gamma\lambda(\alpha^2-1)
+ 9 \lambda^2(1-2\alpha^2))}
{(2 \gamma + 3\lambda)(\gamma-3\lambda)^3} \frac{1}{x^3}
\nonumber  \\
&-& \frac{27 \mu_{\infty} \alpha^2(\lambda-\gamma)}
{(5\gamma+3\lambda)(\gamma-3\lambda)^2}
   \frac{1}{x^4}
+ O(\frac{1}{x^5}) \ .
\end{eqnarray}
It is worthwhile to contrast the coefficient of the $1/x$ correction for the
mass function $\mu(x)$ appearing in  the above equation (\ref{muinfty}) 
with its counterpart in \cite{li}, eq.(18). While in the latter
the coefficient is independent on both the cosmological constant and the
self-coupling of the Goldstone field, we find here a non-trivial dependence
on these parameters (which is indeed confirmed by our numerical analysis).
We also remark that the expansion presented in \cite{li} is in contradiction
with the figure presented in that paper. The figure of \cite{li} seems incompatible
with the fact that the first asymptotic correction to the mass is
supposed to be independent of the cosmological constant.
 
\section{Numerical results}
We remark that without loosing generality, we can choose $\lambda=1.0$.

\subsection{The $\gamma=0$ limit}
This limit was studied previously in great detail in \cite{vile,harari} and \cite{lieb}.
It was found that for $\gamma=0$ 
global monopoles have a negative mass \cite{vile,harari}. Thus, the global monopole
has a repulsive effect on a test particle in its neighbourhood.
Further, it was shown \cite{lieb} that global monopoles without horizon only
exist for $\alpha < \sqrt{\frac{1}{2}}$, while for $\alpha > \sqrt{\frac{3}{2}}$ no static
solutions exist at all. The configurations for $\sqrt{\frac{1}{2}}< \alpha < \sqrt{\frac{3}{2}}$
were called ``supermassive'' monopoles. 

We have redone the calculations and found perfect agreement with the results
in \cite{vile,harari,lieb}. Especially, we remark in view of Fig.~1 that
for $\gamma=0$, the mass of the solution is close to
$-\frac{\pi}{2}$. Since the mass of the global monopole
in flat space is just $=-\frac{\pi}{2}$ \cite{harari} (of course in rescaled units in comparison
to here), the value of $\frac{\mu_{\infty}}{\alpha^2} \lesssim -\frac{\pi}{2}$
is in good agreement with these results. We found in addition that the mass function at large $x$ is always negative
and becomes more and more negative for increasing $\alpha$ \cite{remark1}.

\subsection{Anti-de Sitter (AdS) monopoles}

By solving the three equations numerically we constructed
solutions for negative values of $\gamma$. First, we checked whether the asymptotic behaviour found in (\ref{ainfty})
is correct. We indeed confirmed numerically that the coefficients have the
given dependence on the coupling constants

Following the investigation in \cite{li}, we have then studied the dependence of the mass
$\mu_{\infty}/\alpha^2$ on the cosmological constant $\gamma$ for a fixed value
 of $\alpha$.
As is demonstrated in Fig.~1 for $\alpha=0.1$ and $\alpha=1.0$, we find 
that the mass increases, but stays negative for {\bf all}
values of the cosmological constant. 
In fact, as is evident from Fig.~1, the mass decreases 
further with the increase
of $\alpha$ for a fixed $\gamma$. In the limit $\gamma\rightarrow -\infty$, the
mass tends to zero.
We contrast the behaviour of the mass-function $\mu(x)$ 
for different values of $\gamma$ with that in Fig.1
of \cite{li}. Since the authors choose the vev of the Goldstone field to be $=0.01$, while we choose it to
be $1$, the values of the negative valued cosmological constant in their plot corresponds to choosing
$\gamma=-10$, $-3$ here. Moreover, their choice of $G$ leads to $\alpha=0.0355$.
In Fig.~2, we show the profile of the mass function $\mu(x)$ 
for this choice of parameters. Clearly, the mass
function is negative for all $x$. Of course, we have to add that we have integrated
the equations only up to some maximal value of $x=x_{max}\approx 200$. 
However, as can been seen for the asymptotic expansion (\ref{muinfty}), the
derivative of $\mu(x)$ at large $x$ is always negative and thus the function continues
to decrease. Because of that, local minima or even zeros of the function $\mu(x)$ in the 
asymptotic region are excluded. 

Fixing $\gamma$ and varying $\alpha$, we observe a phenomenon not previously discussed
in the literature. This is demonstrated in Fig.~3 for $\gamma=-0.1$. Increasing $\alpha$, we observe that
a horizon starts to form and at $\alpha=\alpha_{cr}(\gamma)$, the solution
has a degenerate horizon at $x=x_h(\alpha_{cr})$. Thus AdS monopoles without horizon only
exist for $\alpha < \alpha_{cr}$. We find that the value of $\alpha_{cr}$ depends on $\gamma$ and 
that it is increasing with the decrease of $\gamma$. E.g. we find that $\alpha_{cr}(\gamma=-0.1)\approx 1.1$ and
$\alpha_{cr}(\gamma=-1.0)\approx 1.85$ . Note that the solution outside the horizon can {\bf not} be completely described
by a AdS solution of the form:
\begin{equation}
\label{ads}
N(x)=1-2\alpha^2-\frac{2\mu_{\infty}}{x}-\frac{\gamma}{3}x^2
\end{equation}
where $\mu_{\infty}/\alpha^2$ is the mass of the solution. The reason is that $h(x)\equiv 1$ is not
a solution of (\ref{feq}). However, $h(x)$ is close to $1$ for $x > x_h$ and thus (\ref{ads}) can be thought of as 
an approximation.
The solution (\ref{ads}) has a degenerate horizon at $x^{a}_{h}=\sqrt{(1-2\alpha^2)/\gamma}$ with corresponding
mass $\mu_{\infty}^{a}/\alpha^2=\frac{\gamma}{3\alpha^2}\sqrt{((1-2\alpha^2)/\gamma)^3}$. For the values of the parameters
given in Fig.~3 (especially $\alpha=1.098\approx \alpha_{cr}$), we find that $x_h^a = 3.756$ and $\mu_{\infty}^{a}/\alpha^2= -1.465$.
From our numerical analysis, we obtain $x_h\approx 3.5$ and $\mu_{\infty}/\alpha^2\approx -1.7$.
 
Finally, to study the appearance of horizons in this model in more detail,
we have fixed $\alpha=0.7 < \sqrt{1/2}$ and $\alpha=0.8 > \sqrt{1/2}$,
respectively, and studied the dependence of
the value of the zero of $N(x)$, $x_h$ with $N(x_h)=0$
in dependence on $\gamma$. We have chosen these two values of $\alpha$
because only for $\alpha \ge \sqrt{1/2}$ do horizons appear in the
asymptotically flat case $(\gamma=0)$ \cite{lieb}. Our results
are shown in Fig.~4. Clearly, for $\alpha=0.7$ and $\gamma \le 0$ no horizons
appear which confirms the results of \cite{lieb}, while for $\alpha=0.8$ and
$\gamma \le 0$ we find horizons. In fact, for a specific range of $\gamma < 0$
two horizons exist. For $\alpha=0.8$, we find that this is for $-0.00236 \le
\gamma < 0$. At $\gamma=-0.00236$ the two horizons join and form a degenerate
horizon of type shown in Fig.~3. Thus, we have $\alpha_{cr}(-0.00236)=0.8$ which
is in good agreement with the previously presented results. In the limit
$\gamma\rightarrow 0$, the outer horizon tends to infinity, while the inner
horizon tends to the one of the ``supermassive'' monopoles observed
previously \cite{lieb}. 

\subsection{de Sitter monopoles}

In \cite{li} the question was addressed whether the mass of the global monopole
can become positive for specific choices of the
cosmological constant. It was found that for positive cosmological
constants 
this is possible. As a check for a future publication \cite{ybe} on composite 
monopole defects in dS/AdS space-time, we have tried to obtain the results given in \cite{li}
and found contradictions.

Choosing $\gamma$ positive a cosmological horizon appears at $x=x_c(\gamma)$
in dS space. We find that $x_c$ is a decreasing function of the cosmological
constant. For increasing $\gamma$, the value of $x_c=x_0$ tends to zero
as is demonstrated in Fig.~4.
We find further that
$\left( \mu_{\infty}(\gamma > 0)-\mu_{\infty}(\gamma=0)\right)/\alpha^2 < 0$ for all $\gamma >0$.
Since the mass curve does only alter its shape very little when choosing different $\alpha$,
we conclude that in contrast to what is claimed in \cite{li}, the appearance of
a cosmological constant (of either sign) can {\bf not} alter the sign of the mass of the global monopole.
Rather, we find that the mass gets more negative for increasing $\gamma >0$ which can be related to the
fact that the core of the monopole increases due to increased cosmological expansion.

In Fig.~2, we present the mass function for $\alpha=0.0355$ and $\gamma=0.073$.
This should be compared to the Fig.1 in \cite{li}.  
First, we remark that we are surprised that the authors of \cite{li} have managed to find
solutions which correspond to our $\gamma=5$. We find that increasing $\gamma$ from zero to positive values,
we can construct solutions only for $\gamma \lesssim 0.073$. The reason is that with increasing
cosmological constant the horizon which appears for the dS solutions decreases to lie closer and closer to
the core of the monopole. 
Clearly, the mass function $\mu(x)$ is a constantly decreasing function of the coordinate $x$ and
doesn't have local extrema like in \cite{li}. Moreover, the asymptotic values of $\mu(x)$ are always negative.

\section{Conclusions}
Topological defects \cite{vilenkin} are believed to be relevant 
for structure formation in the universe. 
Global defects, i.e. defects which don't involve gauge fields
are of special interest in this context since they have
a long-range scalar field. This leads to the infiniteness of energy
in flat space, but however renders a strong gravitational
effect when the topological defects are studied in curved space.
Moreover, in the case of the global monopole, the coupling to gravity
can remove the singularity present in flat space. The space-time then
has a deficit angle and is not locally flat. Moreover, the mass
of the monopole is negative, which was interpreted as a repulsive effect
of the monopole. 

While for positive cosmological constant (dS space) a horizon, the so-called ``cosmological
horizon'' always appears independent on the gravitational coupling $\alpha$, the existence
of horizons in AdS space depends strongly on the values of the cosmological and gravitational constants.
For vanishing cosmological constant $\gamma=0$, it was found previously that
horizons exist only for $\alpha \ge \sqrt{1/2}$ \cite{lieb}.
For $\gamma < 0$ a monotonically decreasing curve in the $\gamma$-$\alpha$-plane
appears which represents the solutions with one, degenerate horizon.
Above this curve, solutions with two horizons exist, while below the solutions
have no horizons at all.

The authors of \cite{li} have studied global monopoles in a dS/AdS space-time and
found that the inclusion of the cosmological constant can render the mass of the 
monopole positive. Reconsidering these solutions with a highly accurate numerical routine (see 
\cite{bhk} for a short description) and
studying the asymptotic behaviour  we come to a different conclusion~:
global monopoles do {\bf not} acquire a positive mass in AdS or dS space-time.

\begin{acknowledgments}
BH was supported by an EPSRC grant. YB gratefully acknowledges the Belgian FNRS
for financial support.
\end{acknowledgments}

\newpage
\begin{figure}
\centering
\epsfysize=10cm
\mbox{\epsffile{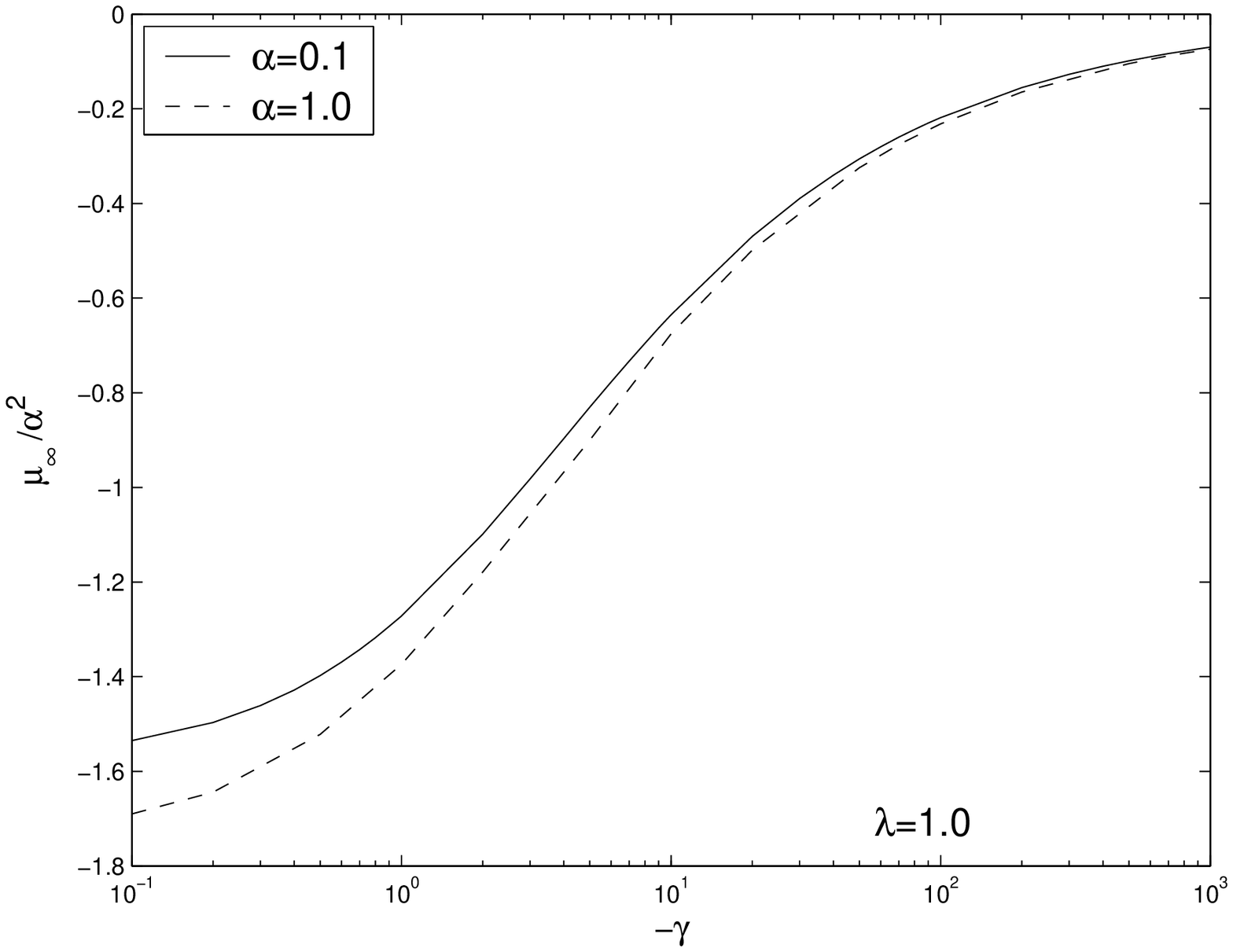}}
\caption{The value of the mass $\mu_{\infty}/\alpha^2$ is given for the AdS monopoles
($\gamma < 0$) as function of $-\gamma$. We have chosen $\alpha=0.1$, =1.0$ and $$\lambda=1.0$. }
\end{figure}
\newpage
\begin{figure}
\centering
\epsfysize=10cm
\mbox{\epsffile{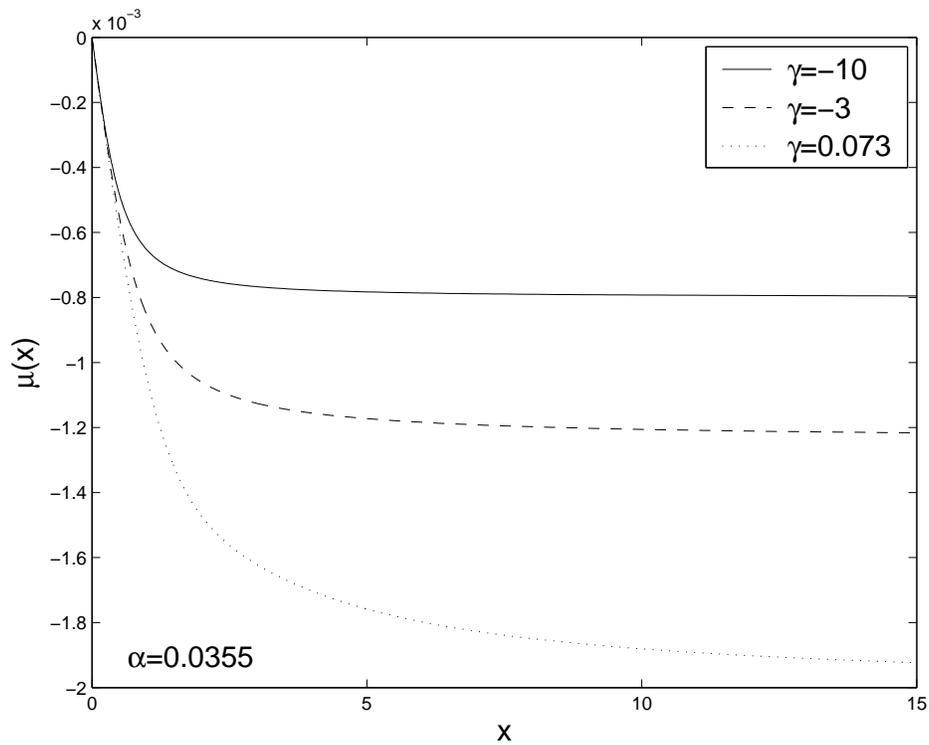}}
\caption{The mass function $\mu(x)$ is shown as function of $x$ for
$\gamma=-10$, $-3$ and $0.073$. We have chosen $\alpha=0.0355$, $\lambda=1.0$. }
\end{figure}

\begin{figure}
\centering
\epsfysize=10cm
\mbox{\epsffile{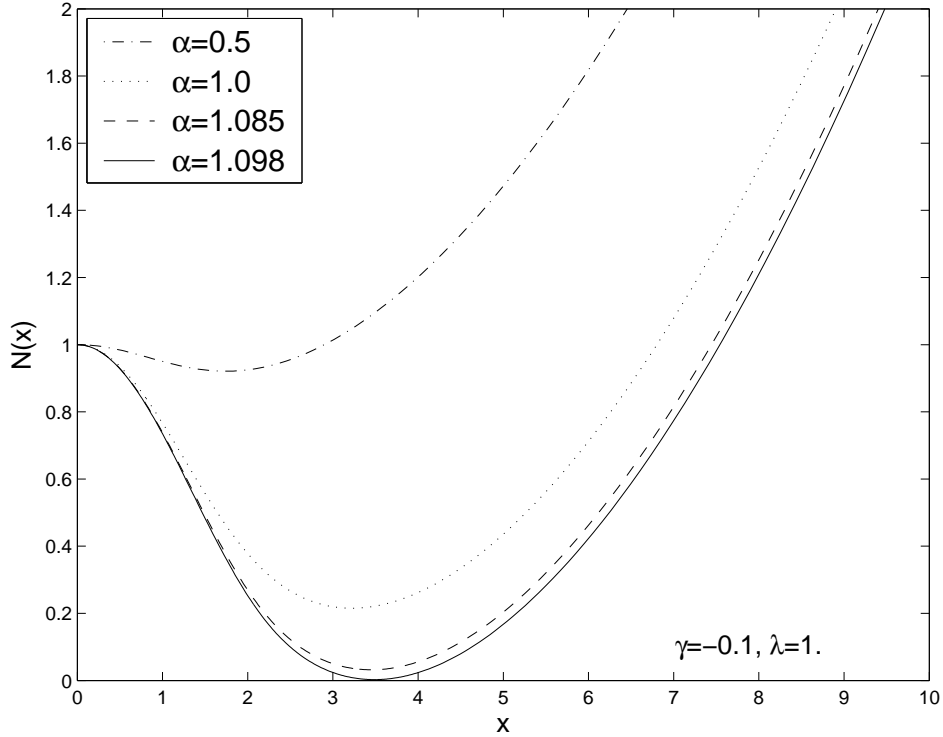}}
\caption{The profile of the metric function $N(x)$ is shown for $\gamma=-0.1$, $\lambda=1.$
and four different choices of $\alpha$, including $\alpha=1.098\approx\alpha_{cr}$.  }
\end{figure}

\begin{figure}
\centering
\epsfysize=10cm
\mbox{\epsffile{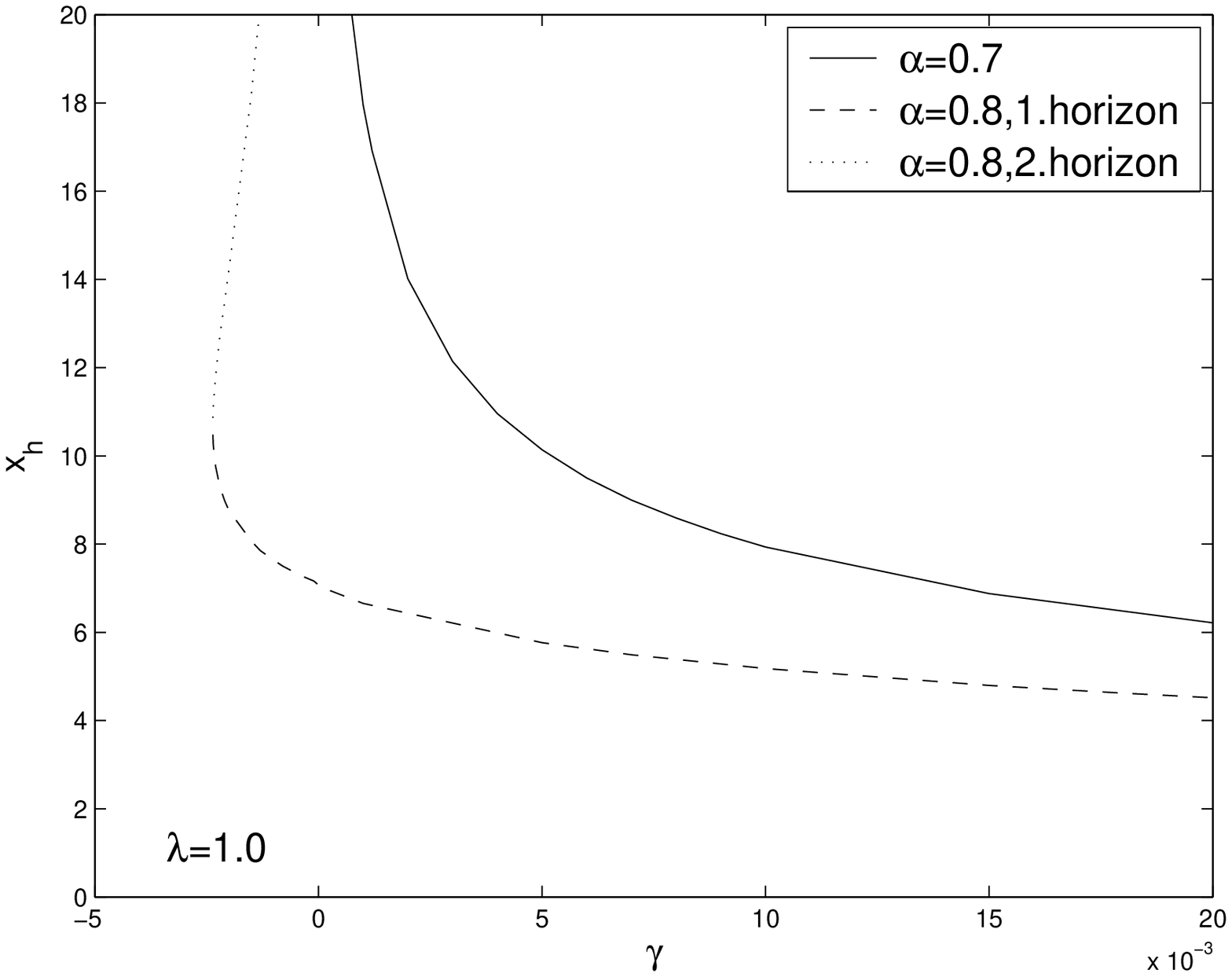}}
\caption{The value of the radial coordinate, where $N(x=x_h)=0$ is shown
as function of the cosmological constant $\gamma$ for two different
values of $\alpha$ and $\lambda=1.0$. We have chosen $\alpha$ smaller (resp. larger) than
$\sqrt{1/2}$, such that for $\gamma=0$ no (resp. one) horizon appears. 
 }
\end{figure}

\begin{thebibliography}{99}
\bibitem{vilenkin}
A. Vilenkin and E. P. S. Shellard, {\it Cosmic Strings and Other
topological defects}, Cambridge University Press, 1994.
\bibitem{strominger}  A. Strominger, JHEP {\bf 0110}, 034 (2001);
JHEP {\bf 0111}, 049 (2001).
\bibitem{adscft} J. Maldacena, Adv. Theor. Phys. {\bf 2}, 231 (1998);
E. Witten, Adv. Theor. Math. Phys. {\bf 2}, 253 (1998).
\bibitem{super} S. Perlmutter et al., Astrophys. J. {\bf 517}, 565 (1999);
A. G. Riess et al., Astron. J. {\bf 116}, 1009 (1998).
\bibitem{vile} M. Bariolla and A. Vilenkin, 
Phys. Rev. Lett. {\bf 63}, 341 (1989).
\bibitem{harari} D. Harari and C. Lousto
Phys. Rev. {\bf D42}, 2626  (1990).
\bibitem{lieb} S. Liebling, Phys. Rev. {\bf D61}, 024030  (1999).
\bibitem{laguna} P. Laguna and D. Garfinkle, Phys. Rev. {\bf D40}, 1011 (1989).
\bibitem{li} X. Li and J. Hao, Phys. Rev. {\bf D66}, 107701 (2002).
\bibitem{com} J. Spinelly, U. de Freitas and E. R. Bezerra de Mello,
Phys. Rev. {\bf D66}, 024018  (2002);
Y. Brihaye and B. Hartmann, Phys. Rev. {\bf D66}, 064018  (2002);
E. R. Bezerra de Mello, Y. Brihaye and B. Hartmann, 
Phys. Rev. {\bf D67}, 045015  (2003).
\bibitem{ybe} E. R. Bezerra de Mello, Y. Brihaye and B. Hartmann, {\it in preparation}
\bibitem{remark1} Note that the mass function $m_L(x)$ used in \cite{lieb} and the function $\mu(x)$
used here are defined differently. It is $m_L(x)=x \alpha^2 + \mu(x)$. Thus even if we find $\mu(x) < 0$
the mass function $m_L(x)$ stays positive asymptotically:
$m_L(x >>1) \sim x \alpha^2 > 0$.
\bibitem{bhk} Y. Brihaye, B. Hartmann and J. Kunz, 
Phys. Rev. {\bf D65}, 024019  (2002).







\end{thebibliography}
\end{document}